\documentclass[12pt,a4paper]{article}
\usepackage{amsmath}
\usepackage[dvips]{graphicx}
\input epsf
\usepackage{times}
\begin{document}
\begin{titlepage}
\title{Mechanism of single-spin asymmetries generation  in the inclusive hadron processes}
\author{S.M. Troshin,
 N.E. Tyurin\\[1ex]
\small  \it Institute for High Energy Physics,\\
\small  \it Protvino, Moscow Region, 142280, Russia} \normalsize
\date{}
\maketitle

\begin{abstract}

We discuss a nonperturbative mechanism for generation of  the
single-spin asymmetries in hadron interactions. It is based on the
chiral quark model combined with unitarity and impact parameter
picture and provides  explanation for the experimental
regularities observed under the measurements of the spin asymmetries.

\end{abstract}
\end{titlepage}
\setcounter{page}{2}

\section*{Introduction}
Studies of the single-spin asymmetries (SSA's) is a sensitive tool
to probe QCD at small and large distances. Experimentally,
significant SSA's were observed in various processes of elastic
scattering, inclusive and exclusive hadron production.


{\it The processes of hadron interactions} are  complicated, there is no proof of
factorization theorem for these processes and it could results from the real absence
of hard and soft parts of interaction factorization in hadronic reactions.
The origin of SSA in these reactions is not clear. Despite
 significant efforts  in theoretical studies devoted to this problem, the phenomenological
success is rather limited;  at the moment there is no comprehensive approach  able to
describe the existing set of   experimental data on polarization, asymmetries,
spin correlations and the unpolarized cross-sections.
Theoretically, there are various approaches to generation of the nonzero SSA but
prevailing number of
 studies of the SSA's  in the field  are based on
{\it assumed} extended factorization in perturbative QCD with considerations of the
Sivers (structure functions) and/or Collins (fragmentation functions) mechanisms
 \cite{sivers,anselm,burk,collins,others} combined sometime with inclusion of
  the
higher twists contributions to the   scattering amplitude of the seemingly to be hard parton subprocess
   \cite{htw,htw1,htw2}.

{\it The decreasing dependence of SSA with $p_T$} --- common feature for the listed above approaches ---
 has not  been observed experimentally. The  experimental
data including the most recent data obtained at RHIC \cite{fpio},
are consistent with a flat transverse momentum dependence at $p_T\geq 1$ GeV/c. Another important
 point is related to the unpolarized inclusive cross-section.
For example, it has also been demonstrated \cite{bsof} that the description of the inclusive cross-section
for $\pi^0$-production, at the energies lower than the RHIC energies also meets difficulties in
the framework of the perturbative QCD.
Deviation from the pQCD scaling is mostly
 noticeable in the forward region where the most significant asymmetry in the $\pi^0$ production
 in $pp_\uparrow \to \pi^0 X$ has also been observed by STAR collaboration at RHIC \cite{fpio} at
$\sqrt{s}= 200$ GeV (in the fragmentation region of the polarized proton).

Of course, more experimental data are needed to perform a conclusive test of various theoretical approaches
and their predictions should be more specified and elaborated for the observables at the hadronic level.
In this connection it should be noted that one
 of the most interesting and persistent  spin phenomenon  is
a very significant polarization of
$\Lambda$--hyperons which has been discovered almost three  decades ago in collisions of
unpolarized hadron beams \cite{newrev}.
It should be also noted , that the asymmetry $A_N=0$ in the neutral pion production
   in the backward  and midrapidity regions \cite{phenix,proza}.
   SSA has also zero value in the $pp_\uparrow\to pX$,
   while $A_N\neq 0$ in the $pp_\uparrow\to nX$ \cite{togawa}
   in the polarized proton fragmentation region.
The approaches based on the assumed pQCD factorization meet in these processes the problems mentioned above.

Thus, it is (more or less depending on the particular personal taste) evident
 that the problems mentioned above
can be related to the illegitimate  use of the methods based on perturbative
expansion, factorization and accounts for higher twists in the region and in the
processes where they actually
 cannot be valid, and it is the
 kinematical region of the modern experiments dealing with rather modest transverse
momenta and energies.
In contrast, it might happen  that
 SSA's  originate from the genuine
nonperturbative sector of QCD (cf. e.g. \cite{spin02}).
Such point of view, i.e. that the polarization has its roots  in the nonperturbative
sector of QCD was widely used in the earlier models  and becomes less isolated one nowadays.
{\it In the nonperturbative sector of QCD}  the two
phenomena,  confinement and chiral symmetry spontaneous breaking  ($\chi$SB)\cite{mnh}
should be reproduced.
The  relevant scales   are characterized by the
parameters $\Lambda _{QCD}$ and $\Lambda _\chi $, respectively.  Chiral $SU(3)_L\times
SU(3)_R$ symmetry is spontaneously  broken  at the distances
between
these two scales.  The $\chi$SB leads
to generation of quark masses and appearance of quark condensates. It describes
transition of current into  constituent quarks.
  Constituent quarks in its turn are  quasiparticles, i.e. they
are a coherent superposition of bare  quarks and their masses
are comparable to  a hadron mass scale.  Therefore
hadron  is often represented as a loosely bounded system of the
constituent quarks.
These observations on the hadron structure lead
to  understanding of several regularities observed in hadron
interactions at large distances. It is well known  that such picture  provides
reasonable  values  for the static characteristics of hadrons, for
 instance, their magnetic moments. The other well known  result
   is  appearance of the Goldstone bosons. It  has been successfully applied
for the  explanation of the nucleon spin structure \cite{cheng} including the most recent results
obtained at JLab \cite{jlab}.

It is necessary  to note that the {\it structure functions} are represented by the distorted
  parton distributions in the impact parameter plane in the polarized case \cite{burk}.
  In this work the   approach
 based on nonperturbative QCD has been used to relate
$\Lambda$-polarization with  large magnitude of the transverse
 flavor dipole moment of the transversely polarized baryons.

The instanton--induced SSA generation   \cite{koch,shur}
 relates those asymmetries to a genuine nonperturbative QCD interaction.
It should be noted that the physics of instantons (cf. e.g. \cite{inst})
can provide microscopic explanation for the $\chi$SB\footnote{We are
grateful to Dmitri Diakonov for the interesting communication on this
matter regarding the polarization phenomena.}.

We discuss here the SSA generation based on chiral
quark model ideas (cf. e.g. \cite{mnh}) and the
filtering spin states related to the account of unitarity in the $s$-channel.
It connects polarization with  asymmetry in the
position (impact parameter) space.
We show that the common features of SSA measurements at RHIC and Tevatron (linear
increase of asymmetry with $x_F$ and flat transverse momentum dependence at $p_T>1$ GeV/c)
    can be reproduced and  described
    in the framework of the  semiclassical picture based
   on  the further development
   of the chiral quark model suggested in \cite{csn} and results of its adaptation
   for the treatment of the polarized and unpolarized inclusive cross-sections.
   The above mentioned data obtained at RHIC \cite{star}
    for the unpolarized inclusive cross-section  can be simultaneously
      described. Consistency  with other new experimental
regularities found at RHIC are discussed as well.

\section{Semiclassical mechanism of SSA generation}

As it was argued
the SSA could originate from the
nonperturbative  QCD and is related to
the mechanism of spontaneous chiral symmetry breaking ($\chi$SB) in QCD \cite{bjorken},
leading to generation of quark masses and appearance of quark condensates.

Thus we consider a
 hadron as an extended object consisting of the valence
constituent quarks located in the central core which is embedded into  a quark
condensate. Collective excitations of the condensate are the Goldstone bosons
and the constituent quarks interact via exchange
of Goldstone bosons \cite{diak}. This interaction is mainly due to a pion field  and has therefore
a spin--flip nature.

At the first stage of hadron interaction common effective
self-consistent field  appears. This field is generated by $\bar{Q}Q$ pairs and
pions interacting with quarks. The time of  generation of the effective field $t_{eff}$
\[
t_{eff}\ll t_{int},
\]
where $t_{int}$ is the total interaction time. This assumption on the almost instantaneous
generation of the effective field obtained some support in the very short thermalization time revealed
in heavy-ion collisions at RHIC \cite{therm}.

Valence constituent quarks   are
 scattered simultaneously (due to strong coupling with Goldstone bosons)
and in a quasi-independent way by this effective strong
 field. Such ideas were  used in the model \cite{csn} which has
been applied to description of elastic scattering and hadron production \cite{mult}.

In the initial state of the reaction $pp_\uparrow\to \pi^0 X$ the proton is polarized
 and can be represented in the simple SU(6) model as  following:
 \begin{equation}\label{pr}
 p_\uparrow=\frac{5}{3}U_\uparrow+\frac{1}{3}U_\downarrow+\frac{1}{3}D_\uparrow+
 \frac{2}{3}D_\downarrow.
\end{equation}
We  exploit the common feature of chiral quark models; namely
the constituent quark $Q_\uparrow$
with transverse spin in up-direction can fluctuate into Goldstone boson and
  another constituent quark $Q'_\downarrow$ with opposite spin direction,
   i. e. perform a spin-flip transition \cite{cheng}:
\begin{equation}\label{trans}
Q_\uparrow\to GB+Q'_\downarrow.
\end{equation}

The $\pi^0$-fluctuations of quarks do not change the quark
 flavor and assuming they have
 equal probabilities in the processes:
\begin{equation}\label{transu}
U_{\uparrow,\downarrow}\to \pi^0+U_{\downarrow,\uparrow}
\quad
\mbox{and}
\quad
D_{\uparrow,\downarrow}\to \pi^0+D_{\downarrow,\uparrow},
\end{equation}
the production of $\pi^0$ by the polarized proton $p_\uparrow$ in this simple $SU(6)$
picture can be regarded as
a result of the fluctuation of the constituent quark $Q_\uparrow$ ($Q=U$ or $D$) in the effective
field into the system $\pi^0+Q_{\downarrow}$ (Fig. 1).
\begin{figure}[h]
\begin{center}
  \resizebox{6cm}{!}{\includegraphics*{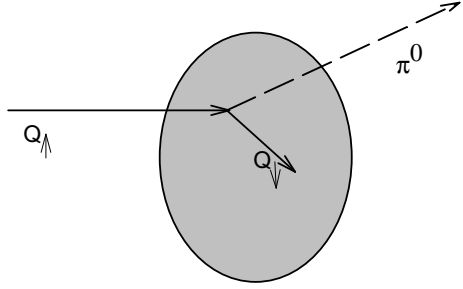}}
\end{center}
\caption{Schematical view of $\pi^0$--production in polarized proton-proton interaction.
 \label{ts1}}
\end{figure}

The contributions to the cross-sections difference
of the quarks polarized in opposite directions compensate each other (as it will be clear in
 what follows),
and it is not the case for the
$\pi^0$-production in the unpolarized case. Therefore the asymmetry $A_N$ should
obey the inequality
$|A_N(\pi^0)|\leq 1/3$.

To compensate quark spin flip $\delta {\bf S}$, an orbital angular momentum
$\delta {\bf L}=-\delta {\bf S}$ should be attributed to the final state of reaction (\ref{trans}).
The presence of $\delta {\bf L}$  in its turn
means  a shift in the impact parameter
value of the Goldstone boson $\pi^0$:
\[
\delta {\bf S}\Rightarrow\delta {\bf L}\Rightarrow\delta\tilde{\bf b}.
\]
Due to   different strengths of interaction at the different
impact distances, i.e.
\begin{eqnarray}
\nonumber p_\uparrow\Rightarrow   Q_\uparrow & \to & \pi^0 + Q_\downarrow\Rightarrow\;
\;-\delta\tilde{\bf b}, \\
\label{spinflip}p_\downarrow\Rightarrow Q_\downarrow & \to & \pi^0 + Q_\uparrow\Rightarrow\;
\;+\delta\tilde {\bf b}.
\end{eqnarray}
the processes of transition $Q_\uparrow$ and $Q_\downarrow$ to $\pi^0$
 will have different probabilities. It eventually leads  to nonzero asymmetry
$A_N(\pi^0)$.
Eqs. (\ref{spinflip}) clarify mechanism of the SSA generation:
when shift in impact
parameter is $-\delta\tilde {\bf b}$ the
interaction is stronger than when the shift is $+\delta\tilde {\bf b}$,
and the asymmetry $A_N(\pi^0)$ is positive.
It is important to note here that the shift in $\tilde{\bf b}$
(the impact parameter of final pion)
is equivalent to the shift of the impact parameter of the initial proton according
to the relation between impact parameters in the multiparticle production\cite{webb}:
\begin{equation}\label{bi}
{\bf b}=\sum_i x_i{ \tilde{\bf  b}_i}.
\end{equation}
The variable $\tilde b$ is conjugated to the transverse momentum of $\pi^0$,
but relations  between functions depending on the impact parameters
$\tilde b_i$,  which will be used further for the calculation of asymmetry,
are nonlinear and therefore we are
using the semiclassical correspondence between small and large values
 of transverse momentum and impact parameter:
\begin{equation}\label{bp}
\mbox{small}\;\tilde b \Leftrightarrow \mbox{large}\;p_T \quad\mbox{and}\quad
\mbox{large}\;\tilde b \Leftrightarrow \mbox{small}\;p_T.
\end{equation}

We consider production of $\pi^0$ in the fragmentation region, i.e.
at large $x_F$ and therefore use the approximate relation
\begin{equation}\label{bx}
b\simeq x_F\tilde b,
\end{equation}
which results from Eq. (\ref{bi}) with  additional assumption on the
small values of Feynman $x_F$ for the other particles. In the symmetrical case of $pp$-interactions
 the model assumes  equal average multiplicities
in the forward and backward hemispheres. It supposes also  small momentum transfer
between the two sides. This is based on the arguments by
 Chou and Yang \cite{yang}.

\section{Asymmetry  and inclusive cross-section}

We apply chiral quark semiclassical mechanism which takes into account
unitarity in the direct channel
to obtain qualitative conclusions
on asymmetry dependence on the kinematical variables.

  The main feature of the mechanism is an account of
unitarity in the direct channel of
reaction. The corresponding formulas for inclusive
cross--sections of the process
\[ h_1 +h_2^\uparrow\rightarrow h_3 +X, \] where hadron $h_3$ in this particular case
 is $\pi^0$ meson and $h_1$, $h_2$ are  protons,
 were obtained in
\cite{tmf} and have the following form
\begin{equation}
{d\sigma^{\uparrow,\downarrow}}/{d\xi}= 8\pi\int_0^\infty
bdb{I^{\uparrow,\downarrow}(s,b,\xi)}/ {|1-iU(s,b)|^2},\label{un}
\end{equation}
where $b$ is the impact  parameter of the initial
protons. Here the function
$U(s,b)$ is the generalized reaction matrix (averaged over initial spin states)
which is determined by the basic dynamics of the elastic scattering.
 The elastic scattering amplitude in the impact
parameter representation $F(s,b)$
   is then given \cite{log}
 by the  relation:
  \begin{equation} F(s,b)=U(s,b)/[1-iU(s,b)].
\label{6} \end{equation}
This equation allows one to obey unitarity provided inequality
  $ \mbox{Im}\,U(s,b)\geq 0\,$  is fulfilled.
The functions $I^{\uparrow,\downarrow}$ in Eq. (\ref{un}) are related   to the
functions   $U_n^{\uparrow,\downarrow}$ --
  the multiparticle
analogs of the function $U$ \cite{tmf} in the polarized case.
The kinematical variables $\xi$
($x_F$ and $p_T$ for example) describe the state of the produced particle
$h_3$.
   Arrows $\uparrow$ and $\downarrow$ denote
   transverse spin directions of the polarized proton $h_2$.

Asymmetry  $A_N$
can be expressed in terms of the functions $I_{-}$, $I_{0}$ and $U$:
\begin{equation} A_N(s,\xi)=\frac{\int_0^\infty bdb
I_-(s,b,\xi)/|1-iU(s,b)|^2} {2\int_0^\infty bdb
I_0(s,b,\xi)/|1-iU(s,b)|^2},\label{xnn}
\end{equation}
where $I_0=1/2(I^\uparrow+I^\downarrow)$ and $I_-=(I^\uparrow-I^\downarrow)$
and $I_0$ obey the sum rule
\[
\int I_0(s,b,\xi) d\xi = \bar n(s,b)Im U(s,b),
\]
here $\bar n(s,b)$ stands for the mean multiplicity in the impact parameter
representation.

On the basis of the described mechanism we can
assume that the functions
$I^\uparrow(s,b,\xi)$ and $I^\downarrow(s,b,\xi)$ are related to the functions
$\frac{1}{3}I_0(s,b,\xi)|_{\tilde b-\delta\tilde b }$ and $\frac{1}{3}I_0(s,b,\xi)|_{\tilde b+
\delta\tilde {b} }$,
respectively, i.e.
\begin{equation}\label{der}
I_-(s,b,\xi)=\frac{1}{3}[I_0(s,b,\xi)|_{\tilde {b}-\delta\tilde {b} }-
I_0(s,b,\xi)|_{\tilde{b}+\delta\tilde{b} }]
=-\frac{2}{3}\frac{\delta I_0(s,b,\xi)}{\delta\tilde{b}}\delta\tilde b.
\end{equation}

We can connect $\delta\tilde b$ with the radius of quark interaction
$r_{Q}^{flip}$
responsible for the transition  changing quark spin:
\[
\delta\tilde b\simeq r_{Q}^{flip}.
\]

Using the above relations and,
in particular, (\ref{bx}), we can write
the following expression for asymmetry $A_N^{\pi^0}$
\begin{equation} A_N^{\pi^0}(s,\xi)\simeq -x_Fr_{Q}^{flip}\frac{1}{3}\frac{\int_0^\infty bdb
I'_0(s,b,\xi)db/|1-iU(s,b)|^2} {\int_0^\infty bdb
I_0(s,b,\xi)/|1-iU(s,b)|^2},\label{poll}
\end{equation}
where $I'_0(s,b,\xi)={dI_0(s,b,\xi)}/{db}$.  In (\ref{poll})
we  made replacement according to relation (\ref{bx}):
\[
{\delta I_0(s,b,\xi)}/{\delta\tilde{b}}\Rightarrow x_F{dI_0(s,b,\xi)}/{db}.
\]
It is clear that $A_N^{\pi^0}(s,\xi)$
(\ref{poll})
should be positive because $I'_0(s,b,\xi)<0$.

The function $U(s,b)$ is
chosen  as a product of the averaged quark amplitudes
in accordance with the quasi-independence of valence constituent
quark scattering in the self-consistent mean field \cite{csn}. The generalized
reaction matrix $U(s,b)$ (in a pure imaginary case, which we consider
 here for simplicity) has
the following form
\begin{equation} U(s,b) = i\tilde U(s,b)=ig(s)\exp(-Mb/\zeta ),
 \label{x}
\end{equation}
where the function $g(s)$ increases at large values of $s$ like a power
of energy:
\[
g(s)= \left[1+\alpha\frac{\sqrt{s}}{m_Q}\right]^N,
\]
$M$ is the total mass of $N$ constituent quarks with mass $m_Q$ in
the initial hadrons and parameter $\zeta$  determines  a universal scale for
the quark interaction radius in the model, i.e. $r_Q=\zeta /m_Q$.

To evaluate asymmetry dependence on $x_F$ and $p_T$
we use semiclassical correspondence  between transverse momentum and impact parameter
  values, Eq. (\ref{bp}).
Performing integration by parts we can rewrite
 the expression for the asymmetry
in the form:
\begin{equation}\label{asym}
A_N^{\pi^0}(s,\xi)\simeq x_Fr_{Q}^{flip}
\frac{M}{3\zeta}\frac{\int_0^\infty bdb
I_0(s,b,\xi)\tilde U(s,b) /[1+\tilde U(s,b)]^3} {\int_0^\infty bdb
I_0(s,b,\xi)/[1+\tilde U(s,b)]^2},
\end{equation}

At small values of $b$  the values of $U$-matrix are large,
and  we can neglect unity in the denominators of the integrands
(however it is rather rough approximation valid only at  high enough
energies).
\begin{figure}[htb]
\begin{center}
  \resizebox{6.5cm}{!}{\includegraphics*{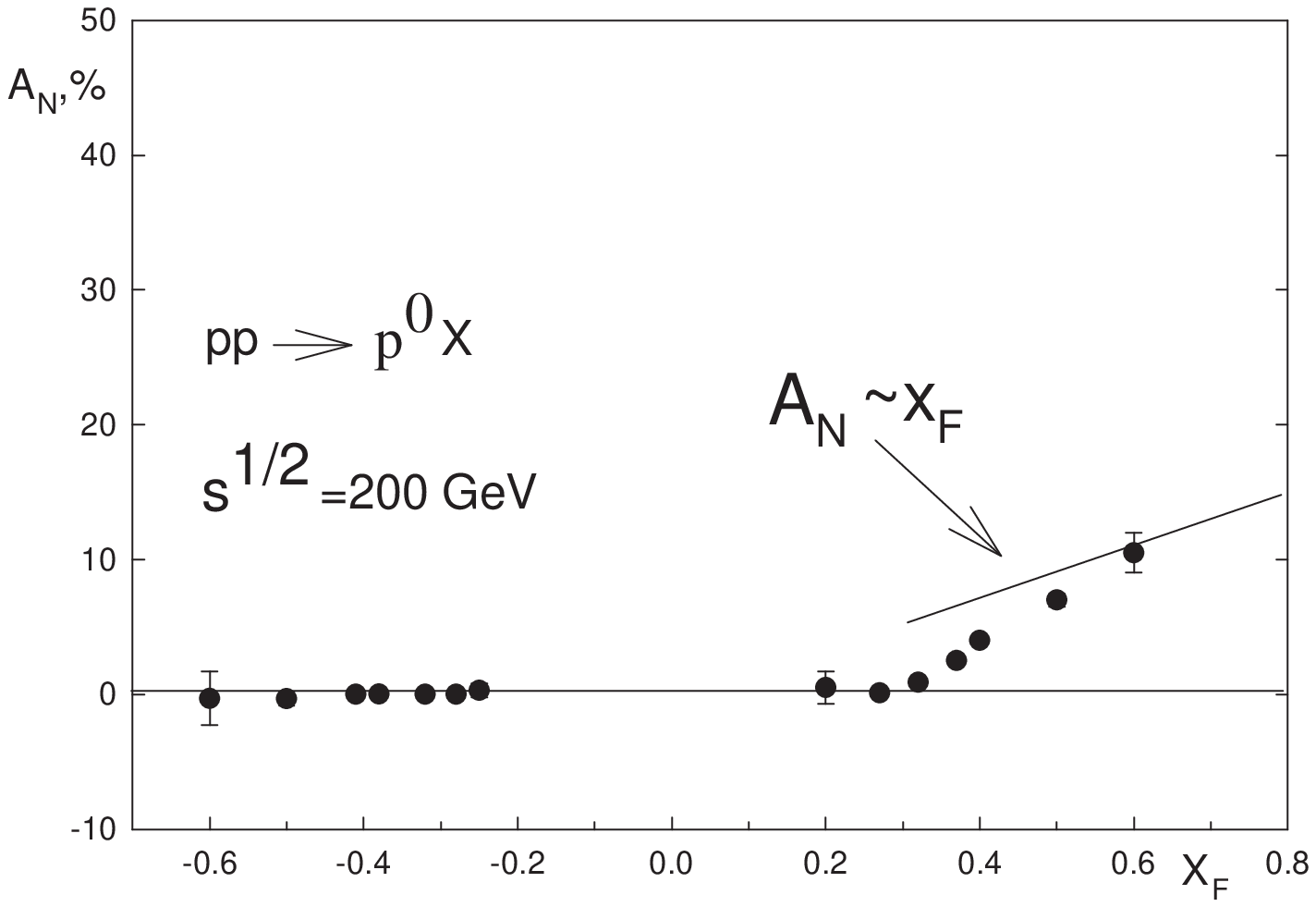}}\;\;\quad
  \resizebox{6.5cm}{!}{\includegraphics*{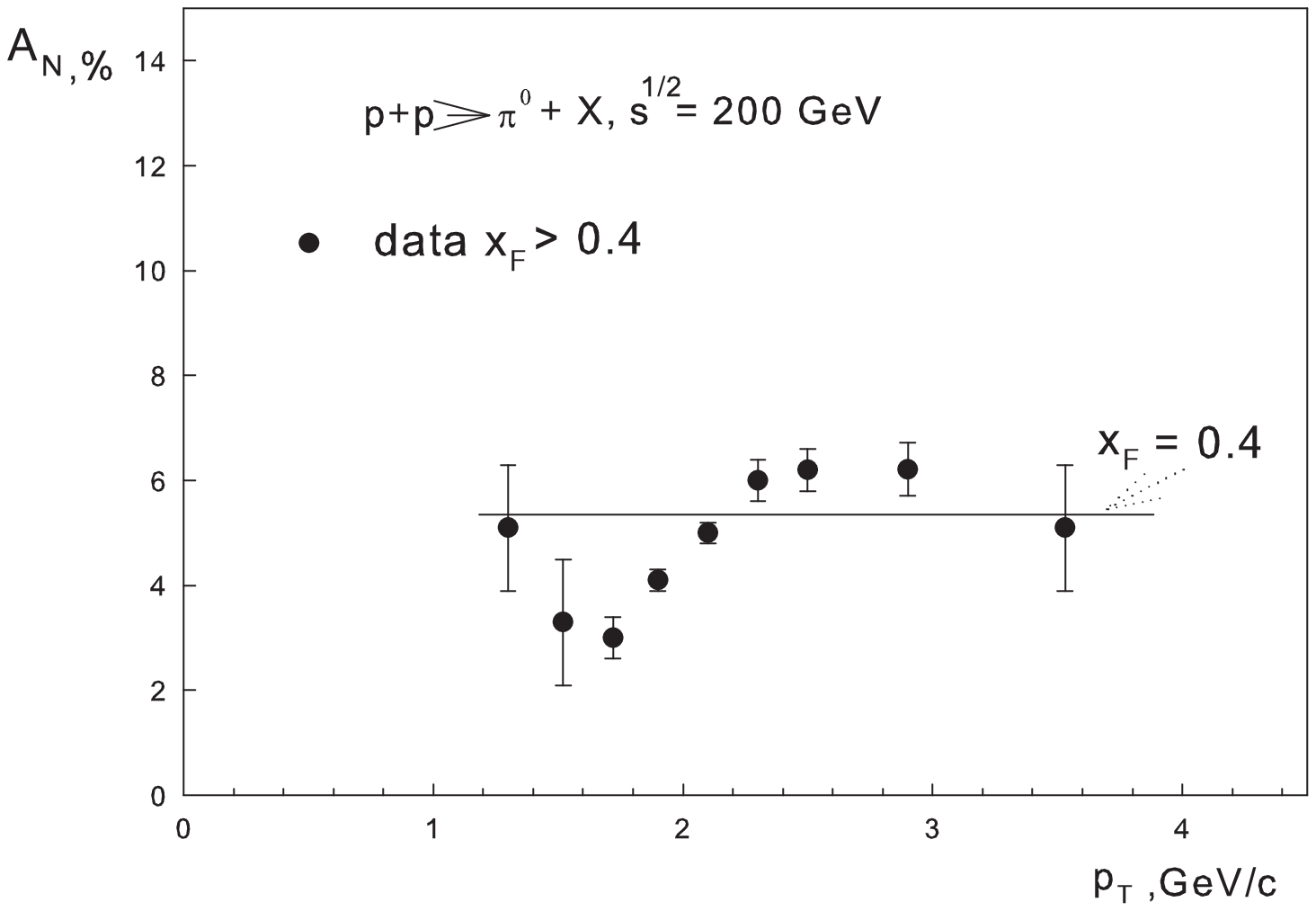}}
\end{center}
\caption{$x_F$ (left panel) and $p_T$ (right panel)
 dependencies of the asymmetry $A_N$ in the process $p+p_\uparrow\to\pi^0+X$ at RHIC,
  experimental data from \cite{fpio}.} \label{ts}
\end{figure}
Thus the ratio of the two integrals (after integration by parts of nominator in Eq. (\ref{asym}))
 is of order
of unity, i.e.  the energy and $p_T$-independent behavior
of asymmetry $A_N^{\pi^0}$ takes place at the values of transverse momentum $p_T\gg x_F/R(s)$:
\begin{equation} A_N^{\pi^0}(s,\xi)\sim x_Fr_{Q}^{flip}
\frac{M}{3\zeta}.\label{polllg}
\end{equation}
Such a  flat transverse momentum dependence of asymmetry results from the similarity of the
rescattering effects for the different spin states, i.e. spin-flip and spin-nonflip
interactions undergo similar absorption at short distances and
the relative magnitude of this absorption does not depend on energy. It is one
of the manifestations of the unitarity.
The numeric value of polarization $A_N^{\pi^0}$ can be significant. Indeed, there is
no small factor in (\ref{polllg}). In Eq. (\ref{polllg}) $M$ is equal
 to the total mass of the constituent quarks in the colliding nucleons,
 the value of parameter $\zeta \simeq 2$.
 We expect that $r_{Q}^{flip}\sim
0.1$ fm on the basis of the model estimate \cite{csn,tmf}.  The above qualitative
 features of asymmetry dependence on $x_F$,
$p_T$ and energy are in  agreement with the experimentally observed trends
 \cite{star}.
For example, Fig. 2 demonstrates that the linear $x_F$ and $p_T$ dependencies is in
 agreement with
the experimental data of STAR Collaboration at RHIC \cite{star}
 in the fragmentation region ($x_F\geq 0.4$). It is this region where the model
should be applicable. Of course,
these dependencies of polarization is the  qualitative ones
and deviations  cannot be excluded.
The same dependencies are compared with the FNAL E704 data \cite{e704} (Fig.3).
Those dependencies  are valid in high-energy
approximation and therefore have been compared with FNAL and RHIC data only. Nevertheless,
they are in a qualitative agreement with the lower energy data also \cite{prz}.
\begin{figure}[htb]
\begin{center}
  \resizebox{6.5cm}{!}{\includegraphics*{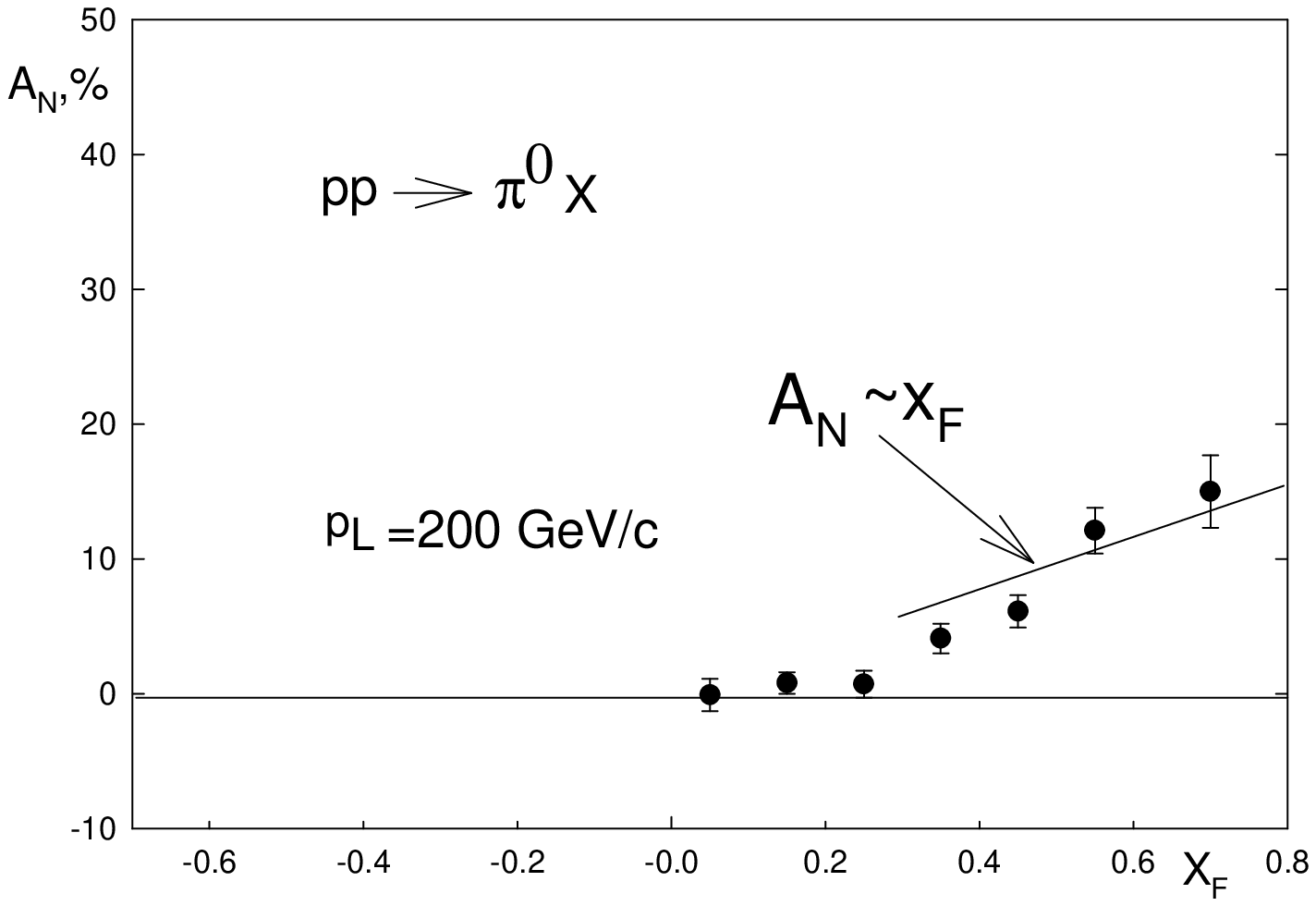}}\;\;\quad
  \resizebox{6.5cm}{!}{\includegraphics*{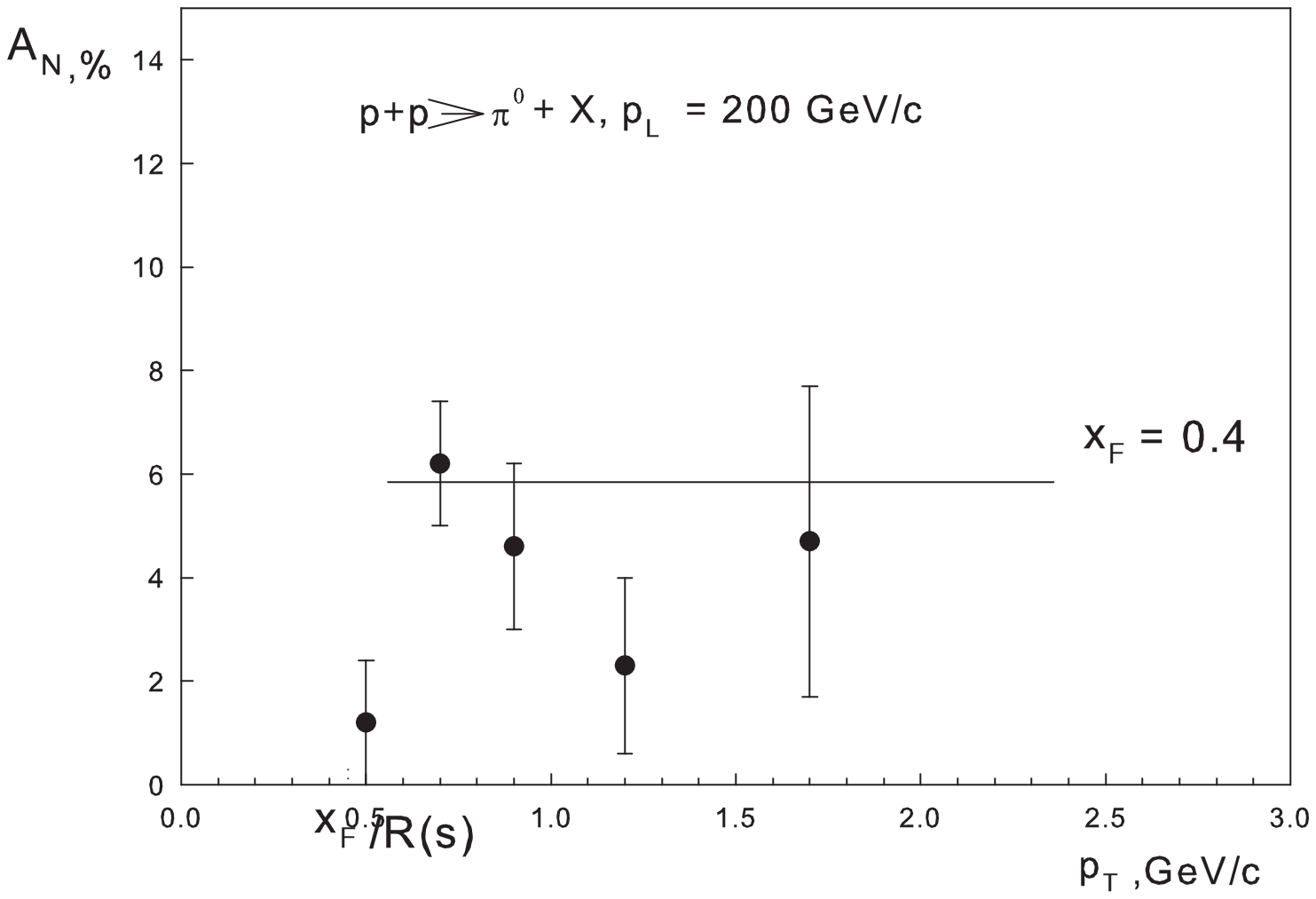}}
\end{center}
\caption{$x_F$ (left panel) and $p_T$ (right panel)
 dependencies of the asymmetry $A_N$ in the process $p+p_\uparrow\to\pi^0+X$ at FNAL,
  experimental data from \cite{e704}.} \label{tsf}
\end{figure}
Comparison with experimental data allows to estimate the value $r_{Q}^{flip}\simeq 0.07$ fm.

Similar mechanism should generate SSA in the production of charged pions.
The relevant process for $\pi^+$--production in polarized $pp_\uparrow$ interactions
\[
U_\uparrow\to\pi^+ + D_\downarrow,
\]
leads to a negative shift in the impact parameter and consequently
to the positive asymmetry $A_N$, while the corresponding process for
the $\pi^-$--production
\[
D_\downarrow\to\pi^- + U_\uparrow
\]
 leads to the positive shift in impact parameter and, respectively,
 to the negative asymmetry $A_N$.
Asymmetry $A_N$ in the $\pi^{\pm}$-production in the
 fragmentation region of polarized proton should have linear $x_F$--dependence
  at $x_F>0.4$ and flat $p_T$ dependence at large $p_T $. Those dependencies
  are similar to the ones depicted on Fig. 2 for $\pi^0$--production. It should be noted
  that at large transverse momenta asymmetries are energy-independent at high energies.

\begin{figure}[h]
\begin{center}
  \resizebox{6cm}{!}{\includegraphics*{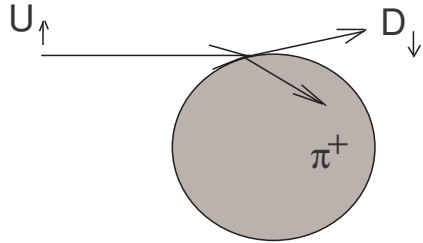}}
\end{center}
\caption{Schematical view of the constituent quark transition on the boundary of effective field.
 \label{ts2}}
\end{figure}

Choosing the region of  small $p_T$ we  select  then the large values of impact parameter
 and  obtain
\begin{equation} A_N^{\pi^0}(s,\xi)\sim x_Fr_{Q}^{flip}
\frac{M}{3\zeta}\frac{\int_{b>R(s)} bdb
I_0(s,b,\xi)\tilde U(s,b) } {\int_{b>R(s)} bdb
I_0(s,b,\xi)},\label{pollsm}
\end{equation}
where $R(s)\sim \ln s$ is the hadron interaction radius, which serve as a scale
separating large and small impact parameter regions.
Eq. (\ref{pollsm}) does not allow to draw a definite conclusion on asymmetry
behaviour. Its dependence   relies  on  the unknown function $I_0(s,b,\xi)$. Nevertheless, it would be interesting
to have at least qualitative estimates of the size of single-spin asymmetries in the
small  momentum transferred region. It should be noted in this connection, that this region includes
the interactions at the boundary
of the effective field localization domain (cf. Fig. \ref{ts2}).
Therefore, in principle, the asymmetry, which is determined
by the variation $\delta I_0(s,b,\xi)/\delta\tilde{b}$, could have significant values
 due
to a large gradient of the interaction intensity in the boundary region. Thus,
 at the values of transverse momenta corresponding
to the constituent quark transition  one can observe very significant asymmetries as
it really happens in the forward neutron production at RHIC (cf. e.g. \cite{phenix}).
Unfortunately, we can not provide the quantitative
estimates of the intensity interaction gradient.
We can  just point out to the possibility, that the
similar phenomena should takes place in other reactions such as polarization of $\Lambda$-hyperons and
it would be important therefore to scan experimentally the region of small transverse momenta in the forward production
by measurements in narrow bins of transverse momentum. It should be noted in this
connection that the chiral quark fluctuation  in the effective field
 with spin flip is
 relatively suppressed when compared to direct elastic
 scattering of quarks  and therefore
does not play a role e.g. in the reaction $pp_\uparrow\to pX$ in the fragmentation
 region, but it should not be suppressed in $pp_\uparrow\to nX$. These features really
 can be observed experimentally: asymmetry $A_N$ is consistent with zero for proton production
 and significantly deviates from zero for neutron production in the forward region.

To underline the model self-consistency we will demonstrate that it is
able to describe
the  unpolarized cross-section of $\pi^0$-production also (Section 5).

\section{Spin filtering and the hyperon polarization}

Chiral quark spin filtering can be used for
the explanation of the hyperon polarization \cite{hypern}. Note that
polarization of $\Lambda$ -- hyperon has the same generic
 dependence on $x_F$ and $p_T$ as the asymmetries in the pion production.
In this section we consider the origin of the hyperon polarization
in the processes where particle in the initial state  are unpolarized.

Experimentally the process
of $\Lambda$-production has been studied more extensively than other hyperon
production processes.
Observed pattern of hyperon polarization is well known and being stable for a long time\footnote{
Polarization of $\Lambda$-hyperons produced in the unpolarized inclusive $pp$--interactions
is negative and energy
independent, it increases linearly with $x_F$ at large transverse momenta
($p_T\geq 1$ GeV/c),
and for such values of transverse momenta   is almost
$p_T$-independent \cite{newrev}.}.

The main idea  is the  filtering or discrimination between
the two initial spin states of equal probability due to different strength of interactions
in the course of scattering in the effective field. The description
 of spin filtering is considered  on the basis of chiral quark model,
formulas for inclusive cross section (with account for the unitarity) \cite{tmf} and
notion on the quasi-independent nature of valence quark scattering in the effective field.

We will use the already discussed feature of chiral quark model that constituent quark $Q_\uparrow$
with transverse spin in up-direction can fluctuate into Goldstone boson and
  another constituent quark $Q'_\downarrow$ with opposite spin direction,
   i. e. perform a spin-flip transition:
\begin{equation}\label{transn}
Q_\uparrow\to GB+Q'_\downarrow\to Q+\bar Q'+Q'_\downarrow.
\end{equation}

To compensate quark spin flip an orbital angular momentum
 should be generated in final state of reaction (\ref{transn}).
The presence of this orbital momentum $\delta {\bf L}$  in its turn
means  shift in the impact parameter
value of the final quark $Q'_\downarrow$ (which is transmitted to the shift in the impact
parameter of $\Lambda$)
\[
\delta {\bf S}\Rightarrow\delta {\bf L}\Rightarrow\delta\tilde{\bf b}.
\]
Due to   different strengths of interaction at the different values of the
impact parameter, the processes of transition to the
spin up and down states will have different probabilities which will lead eventually to
polarization of $\Lambda$.

\begin{figure}[h]
\begin{center}
  \resizebox{4cm}{!}{\includegraphics*{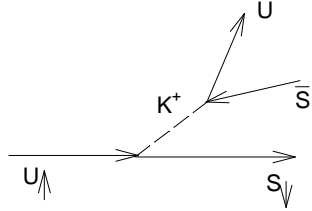}}
\end{center}
\caption{Transition of the spin-up constituent quark $U$ to the spin-down strange quark.
 \label{ts3}}
\end{figure}
In a particular case of $\Lambda$--polarization the relevant transitions
of constituent quark $U$ (cf. Fig. 3) will be correlated with the shifts $\delta\tilde b$
in impact parameter $\tilde b$ of the final
$\Lambda$-hyperon, i.e.:
\begin{eqnarray}
  \nonumber U_\uparrow & \to & K^+ + S_\downarrow\Rightarrow\;\;-\delta\tilde{\bf b} \\
\label{spinflipn} U_\downarrow & \to & K^+ + S_\uparrow\Rightarrow\;\;+\delta\tilde {\bf b}.
\end{eqnarray}
Eqs. (\ref{spinflipn}) clarify mechanism of the   spin states filtering:
 when shift in impact
parameter is $-\delta\tilde {\bf b}$ the
interaction is stronger compared to the case when shift is $+\delta\tilde {\bf b}$,
and the final $S$-quark
(and $\Lambda$-hyperon) is polarized negatively.

 The shift of $\tilde{\bf b}$
(the impact parameter of final hyperon)
is translated then to the shift of the impact parameter of the initial particles.

The mechanism of the polarization generation  is quite natural
 and it has  analogy in optics with the passing
 of the unpolarized light through the glass of polaroid.
Spin states filtering is related to
 emission of Goldstone bosons by constituent quarks.

Now we will  obtain an expression for the polarization which takes into account
unitarity in the direct channel  and apply chiral quark filtering to conclude
on polarization dependence on  kinematical variables.

\section{$\Lambda$-polarization dependence on kinematical variables}
We use the explicit formulas for inclusive
cross--sections of the process
\[ h_1 +h_2\rightarrow h_3^\uparrow +X, \] where hadron $h_3$ is a hyperon whose
transverse polarization is measured, obtained in
\cite{tmf}.  Calculation of polarization of $\Lambda$ proceeds the same steps
as those described in Section 1, i.e.

Polarization  \[ P=
\{\frac{d\sigma^\uparrow}{d\xi}-\frac{d\sigma^\downarrow}{d\xi}\}/
\{\frac{d\sigma^\uparrow}{d\xi}+\frac{d\sigma^\downarrow}{d\xi}\} \]
can be expressed in terms of the functions $I_{-}$, $I_{0}$ and $U$:
\begin{equation} P(s,\xi)=\frac{\int_0^\infty bdb
I_-(s,b,\xi)/|1-iU(s,b)|^2} {2\int_0^\infty bdb
I_0(s,b,\xi)/|1-iU(s,b)|^2},\label{xnnn}
\end{equation}
where $I_0=1/2(I^\uparrow+I^\downarrow)$ and $I_-=(I^\uparrow-I^\downarrow)$.

We can connect $\delta\tilde b$ with the radius of quark interaction
$r_{U\to S}^{flip}$
responsible for the transition $U_\uparrow\to S_\downarrow$ changing quark spin and flavor:
\[
\delta\tilde b\simeq r_{U\to S}^{flip}.
\]

Using the  formulas from previous sections, we will arrive to
the energy and $p_T$-independent behavior
of polarization $P_\Lambda$ at small values of $b$ (and large $p_T$):
\begin{equation} P_\Lambda(s,\xi)\propto -x_Fr_{U\to S}^{flip}
{M}/\zeta.\label{polllgn}
\end{equation}
A numeric value of polarization $P_\Lambda$ can be large: there are again
no small factors in (\ref{polllgn}).
\begin{figure}[htb]
\begin{center}
  \resizebox{6cm}{!}{\includegraphics*{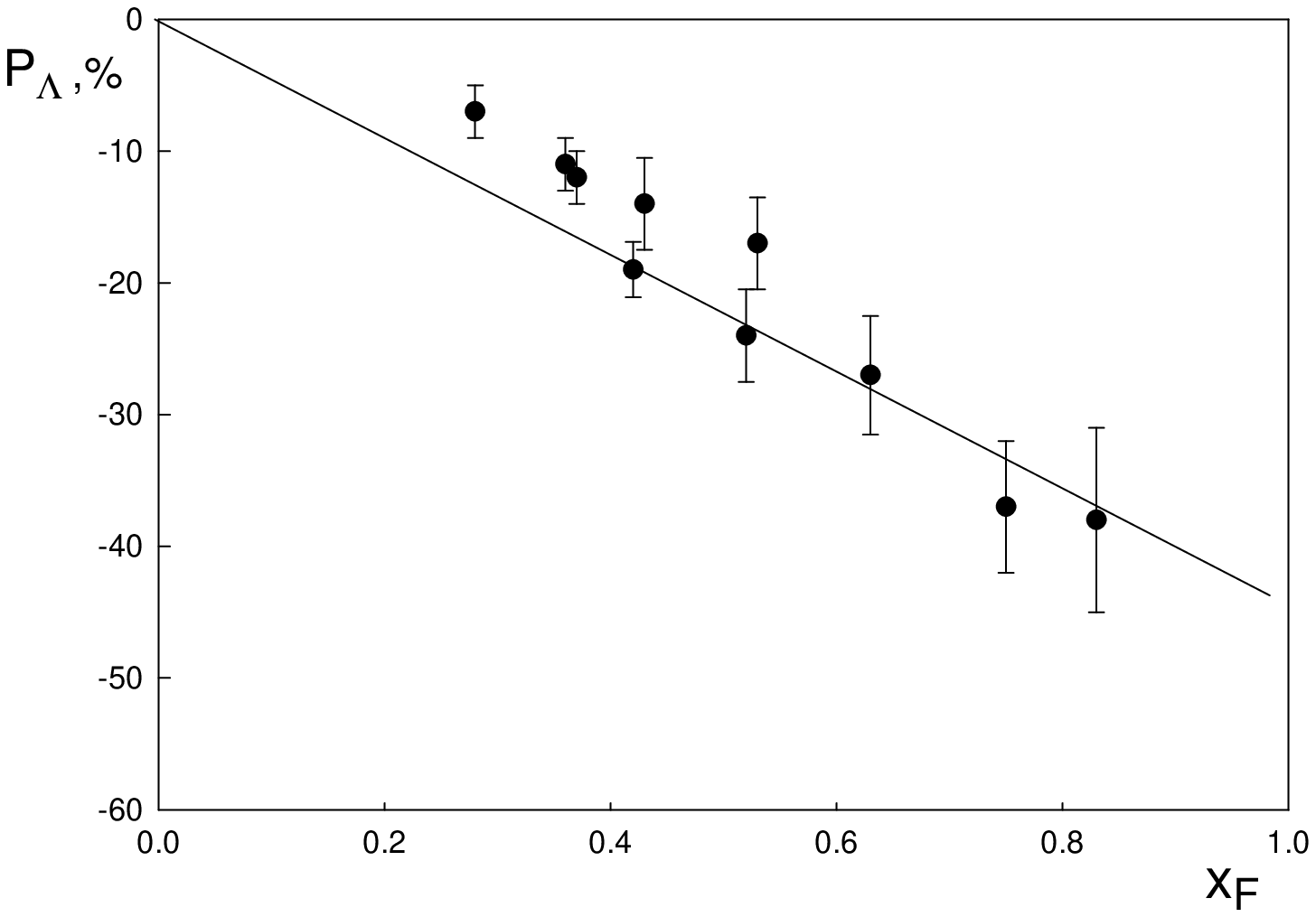}}\;\;\quad
  \resizebox{6cm}{!}{\includegraphics*{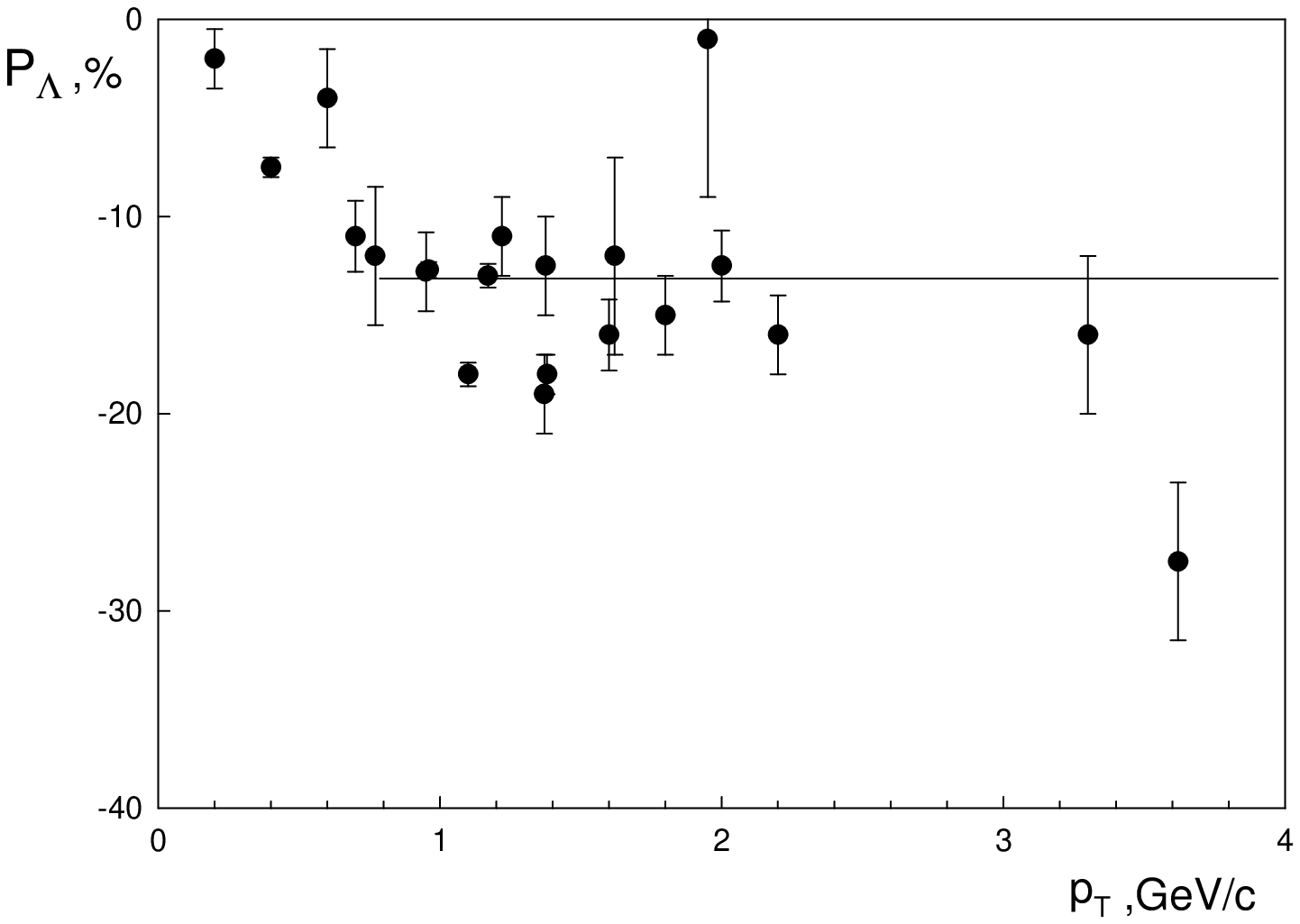}}
\end{center}
\caption{$x_F$ (left panel) and $p_T$ (right panel)
 dependencies of the $\Lambda$-hyperon
polarization} \label{tsr}
\end{figure}
The above qualitative
 features of polarization dependence on $x_F$,
$p_T$ and energy are in a good agreement with the experimentally observed trends
 \cite{newrev}.
For example, Fig. 5 demonstrates that the linear $x_F$ dependence is in a good agreement with
the experimental data in the fragmentation region ($x_F\geq 0.4$) where the model
should work. Of course,
the conclusion on the $p_T$--independence of polarization is a rather approximate one
and deviation from such behavior cannot be excluded.

At small transverse momenta we can write
the following expression for polarization $P_\Lambda(s,\xi)$
\begin{equation} P_\Lambda(s,\xi)\propto -x_Fr_{U\to S}^{flip}
\frac{M}{\zeta}\frac{\int_{b>R(s)} bdb
I_0(s,b,\xi)\tilde U(s,b) } {\int_{b>R(s)} bdb
I_0(s,b,\xi)},\label{pollsmn}
\end{equation}
where $R(s)\propto \ln s$ is the hadron interaction radius, which serve as a scale
of large and small impact parameter values. Polarization dependence in this region
is determined by the unknown function $I_0(s,b,\xi)$ and can have significant values
at the transverse momentum which correspond to scattering in the boundary region of the
effective field.

\section{Inclusive cross-sections of the unpolarized hadron
production}
To demonstrate self-consistency of the model we consider  in this section
the  unpolarized cross-section of $\Lambda$ and pion production processes:
\begin{equation}
\frac{d\sigma}{d\xi}= 8\pi\int_0^\infty
bdb\frac{I_0(s,b,\xi)}{|1-iU(s,b)|^2}\label{unp}.
\end{equation}

At the beginning we approach the process of $\Lambda$-production.
In the fragmentation region we can simplify the problem and consider
the process of $\Lambda$-production as a quasi two-particle reaction,
where the second final particle has a mass $M^2\simeq (1-x_F)s$.
 The amplitude of this quasi two-particle reaction in the pure imaginary
 case (which we consider for simplicity) can be written in the form
 \begin{equation}
F(s,p_T,x_F)= \frac{is}{x_F^2\pi^2}\int_0^\infty
bdbJ_0(bp_T/x_F)\frac{I^{1/2}_0(s,b,x_F)}{1+U(s,b)}\label{amp}.
\end{equation}
To obtain Eq. \ref{amp} we have used relations $b\simeq x_F\tilde{b}$ and due to
the fact that the functions $I_0$ is   quadratic on the the multiparticle
analog of the generalazed reaction matrix $U$ the relation
\begin{equation}\label{img}
I_0^{1/2}(s,b,p_T, x_F)=
\frac{s}{\pi^2}\int_0^\infty I_0^{1/2}(s,b,\tilde{b},x_F)J_0(\tilde{b}p_T)\tilde{b}d\tilde{b}.
\end{equation}

Since in the model constituent quarks are considered to form a $SU(6)$ wave function,
$I_0=I_0^{U\to S}$.
The function $I_0^{U\to S}(s,b,x_F)$ according to quasi-independent nature
of constituent quark-scattering
can be represented then as a product
\begin{equation}
 I_0^{U\to S}(s,b,x_F)= \left[\prod^{N-1}_{Q=1} \langle f_Q(s,b)\rangle\right]\langle
f_{U\to S}(s,b,x_F)\rangle,
\end{equation}
 where $N$ is the total number of quarks in the colliding
hadrons.

In the model the $b$--dependencies of the amplitudes $\langle f_{Q}
\rangle $ and $\langle f_{U\to S} \rangle $
are related to the strong
formfactor of the constituent quark and transitional spin-flip formfactor
 respectively.
 The strong interaction radius of constituent
quark is determined by its mass. We suppose that the corresponding radius of transitional formfactor
 is determined by the average mass $\tilde{m}_Q=(m_U+m_S)/2$ and factor $\kappa<1$ (which takes
into account reduction of the radius due to spin flip)
$r^{flip}_{U\to S} = \kappa\zeta /\tilde{m}_{{Q}}$ and the corresponding function $f_{U\to S}(s,b,x_F)$ has the form
\begin{equation}\label{ftr}
f_{U\to S}(s,b,x_F)=g_{flip}(x_F)\exp\left(-\frac{\tilde{m}_Q}{\kappa\zeta}b\right)
\end{equation}

 The expression for
$I_0(s,b,x_F)$ can be rewritten then in the following form:
\begin{equation}\label{iol}
I_0(s,b,x_F) =\frac{\bar{g}(x_F)}{g_Q(s)}
U(s,b)\exp[-\Delta m_Q b/\zeta ],
\end{equation}
where the mass difference $\Delta m_Q\equiv\tilde{m}_Q/\kappa-m_Q$ and $\bar{g}(x_F)$ is the function whose
dependence  on Feynman $x_F$ in the model is not fixed.

Now we can consider $p_T$- and $x_F$-dependencies
of the $\Lambda$-hyperon production
cross-section and we start with angular distribution\footnote{One should remember that
all formulas and figures below are valid for the fragmentation region only, i.e. for $x_F>0.4$}.
The corresponding amplitude
$F(s,p_T,x_F)$ can be calculated analytically. To do so
 we continue the amplitudes
$F(s,\beta, x_F),\,\beta =b^2$, where
\[
F(s,\beta, x_F)=\frac{1}{x_F^2}\frac{I^{1/2}_0(s,\beta,x_F)}{1+U(s,\beta)}
\]
to the complex
 $\beta $--plane and transform the Fourier--Bessel integral over impact
parameter into the integral in the complex $\beta $ -- plane over
the contour $C$ which goes around the positive semiaxis.
 The amplitude $F(s,\beta, x_F)$ has the poles and a
  branching point (at $\beta =0$) and
therefore the  amplitude $F(s,p_T,x_F)$  can be represented as
a sum of the poles contribution and the contribution of the cut:
\begin{equation}
F(s,p_T,x_F)=F_p(s,p_T,x_F)+F_c(s,p_T,x_F)\label{sum}
\end{equation}
Calculation of poles and cut contributions are similar to the case of
elastic scattering \cite{ech}.

The poles and cut contributions determine the
the inclusive cross-section behaviour of $\Lambda$ production at moderate and large
values of $p_T$ correspondingly, i.e. it will have in
the region of large $p_T$ power-like dependence on $p_T$:
\begin{equation}\label{dsig}
\frac {d\sigma}{d\xi}\propto G_c^2(s,x_F)(1+\frac{p_T^2}{ x^2_F \bar{M}^2})^{-3},
\end{equation}
while at smaller values of $p_T$ inclusive cross-section would have the exponential $p_T$-dependence:
\begin{equation}\label{dsig1}
\frac {d\sigma}{d\xi}\propto G_p^2(s,x_F)\exp(-\frac{2\pi \zeta }{M}\frac{p_T}{x_F}).
\end{equation}
The data for the $\Lambda$-hyperon production are available at
the moderate values of $p_T$ and the experimental fits to the
data \cite{data} of the  form \[
A(1-x_F)^n e^{-B(x_F)p_T}\]
 just
follow to Eq. (\ref{dsig1}) when relevant parameterization for the
function $\bar g(x_F)$ is chosen. At high values of $p_T$ power-like dependence
should take place according to Eq. \ref{dsig}.  In the energy region of $\sqrt{s}\leq 2$ TeV
the functions $G_p$ and $G_c$ have very slow variation with energy due to the numerical values
of parameters \cite{preas}.

We can treat the inclusive cross-section of the pion production processes in a similar way.
In the fragmentation region
at small $p_T$ the poles in impact parameter plane at $b\sim R(s)$
lead to the exponential $p_T$--dependence of inclusive cross-section.
\begin{figure}[htb]
\begin{center}
  \resizebox{8cm}{!}{\includegraphics*{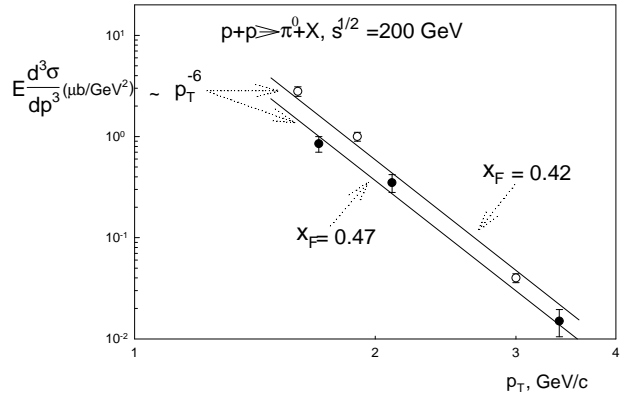}}
\end{center}
\caption{Transverse momentum dependence of unpolarized inclusive cross--section,
experimental data from \cite{star}.} \label{ds}
\end{figure}
At high $p_T$ the power-like dependence $p_T^{-n}$ with $n=6$
should take place. It originates from the singularity at zero impact parameter
 $b=0$.
 The exponent $n$ does not depend
 on $x_F$. The data are in a good agreement with the $p_T^{-6}$--dependence of
the unpolarized inclusive cross--section (Fig. 6). Recently a similar $p_T^{-6}$--dependence
has been obtained for the  soft contribution to quark-quark scattering induced by an anomalous
 chromomagnetic interaction due to instanton mechanism \cite{kpt6}.

Thus, in the approach with effective degrees of freedom --
 constituent quarks and Goldstone bosons -- differential cross--section at high transverse momenta
 has a generic power-like dependencies on $p_T$.
 domain.

\section*{Conclusion}
The considered mechanism of SSA generation deals with the
 effective degrees of freedom and takes into
account collective aspects of QCD dynamics. Combined with unitarity, which is an essential
part of  the approach, it allows to get a qualitative explanation of the observed
 regularities:
linear dependence on $x_F $ and
flat dependence  on transverse momentum at large $p_T$ of SSA's
in the polarized proton fragmentation region.
The  spin filtering is responsible for the generation of hyperon polarization
in the collisions of the unpolarized nucleons. In particular it leads to the similar
behaviour of $\Lambda$-polarization.
The
application
of spin filtering to other hyperons is a more complicated
case,
since those hyperons
 have two or three strange quarks and the spins of  $U$ and $D$
quarks  also make contributions into their polarizations.

We also discussed here  particle production in the fragmentation region and have
shown that the power-like behavior of the differential cross-sections at large transverse
momenta can be obtained in the approach which has a nonperturbative origin.
It is no need to comment that such a dependence
  always being considered as a manifestation of the genuine  hard, short distance
 processes where asymptotic freedom is at work. Power-like  behavior of inclusive cross-sections
  and the strongly
 interacting nature of quark matter revealed at RHIC, in principle, can be attributed to a different dynamics.
 However, it is difficult to
  imagine how the both phenomena can coexist in the strongly interacting coherent medium
  observed at RHIC when thermalization occurs at very early stage of reaction.
  It seems
 natural to suppose that
 they should have the same origin. One  should
 arrive then to conclusion
 that the power-like
 dependence of the differential cross-sections should not necessarily be associated with the
 processes treated by  perturbative QCD. This viewpoint gets support
 in the results on polarization measurements which also indicate possibility of power-like behavior
 due to soft dynamics. It should also be recollected that the energies
 where power-like dependence in exclusive
 processes was  observed are evidently
  too low to be considered as a true asymptotic perturbative QCD. This regime should occur at much higher
  values of the transverse momenta and energy.

Finally, one should note that in the central and backward regions where correlations
 between impact parameters of the initial and  final particles
 are weak or even completely degraded, the asymmetry cannot be generated
 due to the considered  mechanism. The experimentally observed  vanishing asymmetries
 in the central and backward regions  provide  indirect
evidence  in its favor.

\section*{Acknowledgement}
We are grateful to C.~Aidala, A.~Bazilevsky, V.~Mochalov, S.~Shimanskiy,  A.~Vasiliev
and A.~Zelenski for the information and
interesting discussions of the experimental data with their phenomenological interpretations.
\end{document}